\documentclass[journal,onecolumn]{IEEEtran}
\usepackage{cite}
\ifCLASSINFOpdf
\else
\fi
\usepackage[cmex10]{amsmath}
\usepackage{amssymb}
\usepackage{xcolor}

\newcommand{\LSMF}{L_\text{SMF}}
\newcommand{\NSMF}{{N_\text{SMF}}}

\newcommand{\expval}{\operatorname{\mathbb{E}}}

\newcommand{\fourier}{\mathcal{F}}

\newcommand{\rmH}{\mathrm{H}}
\newcommand{\rmT}{\mathrm{T}}

\newcommand{\Ax}{A_x}
\newcommand{\Ay}{A_y}
\newcommand{\Axl}{A_{x,l}}
\newcommand{\Ayl}{A_{y,l}}
\newcommand{\Axp}{A_{x,p}}
\newcommand{\Ayp}{A_{y,p}}
\newcommand{\Axltilde}{\tilde{A}_{x,l}}
\newcommand{\Ayltilde}{\tilde{A}_{y,l}}
\newcommand{\Axptilde}{\tilde{A}_{x,p}}

\newcommand{\Gx}{G_x}
\newcommand{\Gy}{G_y}

\newcommand{\Gxptilde}{G_{x,p}}
\newcommand{\Gyptilde}{G_{y,p}}
\newcommand{\Gxptildea}{G_{x,p,1}}
\newcommand{\Gxptildeb}{G_{x,p,2}}
\newcommand{\Gxptildec}{G_{x,p,3}}
\newcommand{\Gxptilded}{G_{x,p,4}}
\newcommand{\Ra}{R_{1}}
\newcommand{\Rb}{R_{2}}
\newcommand{\Rc}{R_{3}}
\newcommand{\Rd}{R_{4}}

\newcommand{\kxx}{k'\!}
\newcommand{\kx}{k}
\newcommand{\ky}{k}
\newcommand{\lxx}{l'\!}
\newcommand{\lx}{l}
\newcommand{\ly}{l}
\newcommand{\mxx}{m'\!}
\newcommand{\mx}{m}
\newcommand{\my}{m}

\newcommand{\xikx}{\xi_{k}}
\newcommand{\xiky}{\zeta_{k}}
\newcommand{\xilx}{\xi_{l}}
\newcommand{\xily}{\zeta_{l}}
\newcommand{\ximx}{\xi_{m}}
\newcommand{\ximy}{\zeta_{m}}

\newcommand{\xikxx}{\xi_{k'\!}}

\newcommand{\xilxx}{\xi_{l'\!}}
\newcommand{\xilyy}{\zeta_{l'\!}}
\newcommand{\ximxx}{\xi_{m'\!}}
\newcommand{\ximyy}{\zeta_{m'\!}}

\newcommand{\Ltot}{L}

\newcommand{\Cklm}{\mathcal{C}_{klm}}
\newcommand{\Cooo}{\mathcal{C}_{0}}
\newcommand{\Clkm}{\mathcal{C}_{lkm}}
\newcommand{\CCklm}{\mathcal{C}_{k'\!l'\!m'\!}}

\newcommand{\infint}{\int_{-\infty}^{\infty}}

\newcommand{\od}[2]{\frac{d#1}{d#2}}

\newcommand{\pdn}[3]{\frac{\partial^{#3}#1}{\partial#2^{#3}}}
\newcommand{\pd}[2]{\frac{\partial#1}{\partial#2}}

\newcommand{\opt}[1]{}

\begin{document}
\title{Analytical Modeling of Nonlinear Propagation in a Strongly
  Dispersive Optical Communication System}

\author{Pontus Johannisson% <-this % stops a space
\thanks{Pontus Johannisson is with the Photonics Laboratory,
Department of Microtechnology and Nanoscience, Chalmers University of
Technology, SE-412\,96 G\"oteborg, Sweden, e-mail:
pontus.johannisson@chalmers.se.}}


%\markboth{Journal of \LaTeX\ Class Files,~Vol.~6, No.~1, January~2007}%
%{Shell \MakeLowercase{\textit{et al.}}: Bare Demo of IEEEtran.cls for Journals}
\markboth{}%
{}



\maketitle

\begin{abstract}
%\boldmath
Recently an analytical model was presented that treats the nonlinear
signal distortion from the Kerr nonlinearity in optical transmission
systems as additive white Gaussian noise.  This important model
predicts the impact of the Kerr nonlinearity in systems operating at a
high symbol rate and where the accumulated dispersion at the receiver
is large.  Starting from the suggested model for the propagating
signal, we here give an independent and different calculation of the
main result.  The analysis is based on the Manakov equation with
attenuation included and a complete and detailed derivation is given
using a perturbation analysis.  As in the case with the published
model, in addition to assuming that the input signal can be written on
a specific form, two further assumptions are necessary; the
nonlinearity is weak and the signal-noise interaction is neglected.
The result is then found without any further approximations.
\end{abstract}

\begin{IEEEkeywords}
Optical fiber communication, communication system nonlinearities,
nonlinear optics, wavelength division multiplexing.
\end{IEEEkeywords}

% \fi
\IEEEpeerreviewmaketitle

\section{Introduction}




\IEEEPARstart{I}{n an optical} transmission system operating at a high
symbol rate, the propagating signal is rapidly evolving due to
chromatic dispersion (CD).  It has been found numerically that after a
relatively short distance of propagation, the probability density
function of all four quadratures of a polarization-multiplexed signal
become Gaussian with zero mean and a variance related to the signal
power~\cite{carena_2010_ecoc}.  This can also be shown analytically by
using the central limit theorem and is intuitively understandable
since the dispersed signal at every point in time can be viewed as a
coherent superposition of many signal pulses.  During the propagation,
a nonlinear phase shift is induced in proportion to the local power by
the Kerr nonlinearity.  As the CD is compensated in the receiver, this
phase shift will give rise to a residual signal distortion.
Numerically, it has been observed that this distortion is very similar
to additive white Gaussian noise (AWGN)~\cite{carena_2010_ecoc}.  This
observation is of great importance as it suggests that if no attempt
is made to compensate for the nonlinear effects, then modeling the
nonlinear signal distortion as AWGN is possible.

In 2011, an analytical model was presented that calculates the
noise-like nonlinear distortion for a quite general wavelength
division multiplexing (WDM) system~\cite{poggiolini_2011_ptl2,
poggiolini_2011_icton}.  This signal distortion was named
\emph{nonlinear interference} (NLI) and recently an extensive paper on
the same topic was published~\cite{carena_2012_jlt}. The work reported
here is closely related to these publications but it should be pointed
out that there are many results in the literature that address the
same or similar questions.  To name one example, the work for OFDM by
Chen and Shieh~\cite{chen_2010_oe} has many similarities in both the
approach and the results.  However, the previous work in the area
seems to be well described in the introduction
of~\cite{carena_2012_jlt} and we will not further elaborate this topic
here.

In the publications~\cite{poggiolini_2011_ptl2,
poggiolini_2011_icton,carena_2012_jlt}, a suggestion is given for how
to model the signal, the four-wave mixing (FWM) of the different
signal spectral components is calculated, and the corresponding power
spectral density (PSD) is found.  However, the derivation of the main
result is quite short.  Partly this is because known FWM results are
used~\cite{inoue_1995_jlt}.  In this paper, an independent detailed
derivation of the resulting PSD is carried out.  The calculation
starts from the Manakov equation with power gain and attenuation
included and we use the signal model suggested
in~\cite{poggiolini_2011_ptl2}.  The calculation is based on a
perturbation approach previously used, e.g., to investigate
intrachannel cross-phase modulation (XPM) and intrachannel FWM, which
gives rise to ``ghost pulses'' in systems using on-off
keying~\cite{mecozzi_2000_ptl,ablowitz_2000_ol,johannisson_2001_ol}.
This approach has later been used both to analytically study systems,
see for example~\cite{bononi_2008_crp,bononi_2012_oe}, and to
compensate for the NLI, see for example~\cite{yan_2011_ecoc}. Using
the perturbation approach, we here present a self-contained
calculation of the PSD of the NLI directly from the model equation.
This work also goes beyond~\cite{carena_2012_jlt} since, as described
in Section~\ref{sec_formal_solution}, we perform the calculation for a
more general system.

The organization of this paper is as follows: In
Section~\ref{section_pert_ana}, the perturbation analysis is
introduced and a formal solution that is valid for all input signals
is given.  In Section~\ref{section_signal_model}, we find the solution
corresponding to the specific input signal suggested
in~\cite{poggiolini_2011_ptl2}, which is then used in
Section~\ref{section_psd} to calculate the NLI, i.e., the PSD of the
perturbation.  We then calculate the PSD corresponding to the result
in ~\cite[Eq.~(18)]{carena_2012_jlt} by studying a specific system
choice in Section~\ref{section_example}.  Finally, we conclude.

\section{Perturbation analysis}
\label{section_pert_ana}
The calculation of the NLI in~\cite{poggiolini_2011_ptl2} assumes that
the FWM is a weak effect.  This assumption is stated explicitly as
``the pump is undepleted''.  We proceed in a similar way by
introducing the complex envelope of the electric field in the $x$ and
$y$ polarizations according to $\mathbf{A} = (\Ax, \Ay)^\rmT$ and
writing this as $\mathbf{A} = \mathbf{A}_l + \mathbf{A}_p$, where
$\mathbf{A}_l$ is the linearly propagating signal, i.e., the signal in
the absence of any Kerr nonlinearity, and $\mathbf{A}_p$ is a small
perturbation.  This implies that the nonlinear effects may not become
significant and this constitutes the first assumption of the model.
The second assumption is that the input signal can be written on the
form suggested in~\cite{poggiolini_2011_ptl2}.  The range of validity
for this signal model is a separate question and we do not discuss
this question here.  A third assumption is that the signal-noise
interaction is neglected as the calculation of the NLI does not
involve amplifier noise in any way.  This is also an assumption
of~\cite{poggiolini_2011_ptl2} and in order to account for amplifier
noise under this assumption, noise corresponding to the total AWGN
from the amplification is added just before the receiver.  These three
assumptions are sufficient and no further approximations are necessary
to carry out the analysis.

\subsection{The perturbation equation from the Manakov equation}
In order to describe transmission using polarization multiplexing, we
start from the Manakov equation~\cite{wang_1999_jlt} and include power
gain and attenuation.  It should be noted that the Manakov equation is
obtained by averaging over the polarization rotations which are
assumed to be fast and this equation does not take polarization mode
dispersion into account.  Denoting the group-velocity dispersion by
$\beta_2(z)$, the power gain by $g(z)$, and the power attenuation by
$\alpha(z)$ we have
\begin{align}
\label{eq_manakov}
i \pd{\mathbf{A}}{z} = \frac{\beta_2(z)}{2} \pdn{\!\mathbf{A}}{t}{2} -
\gamma(z) (\mathbf{A}^\rmH \mathbf{A}) \mathbf{A} + i \frac{g(z) -
\alpha(z)}{2} \mathbf{A},
\end{align}
where $\mathbf{A}^\rmH \mathbf{A} = |\Ax|^2 + |\Ay|^2$ is the sum of
the power in the $x$ and $y$ polarizations and the nonlinear parameter
$\gamma(z) = (8/9) (k_0 n_2/A_\text{eff})$.  In the expression for
$\gamma$, $k_0$ is the wavenumber corresponding to the center
frequency, $n_2$ is the Kerr coefficient, and $A_\text{eff}$ is the
effective area of the optical fiber.  The power gain $g(z)$, which is
set up using erbium-doped fiber amplifiers (EDFAs) and/or Raman
amplification, is assumed to have no frequency dependence, i.e., the
gain is flat over the bandwidth of the signal.  The perturbation is
introduced according to
\begin{align}
\mathbf{A} = \mathbf{A}_l + \mathbf{A}_p =
\begin{pmatrix}
\Axl \\ \Ayl
\end{pmatrix}
+
\begin{pmatrix}
\Axp \\ \Ayp
\end{pmatrix},
\end{align}
where $\mathbf{A}_l(z, t)$ solves the linear equation obtained by
setting $\gamma = 0$ in (\ref{eq_manakov}).   We find that
\begin{align}
(\mathbf{A}^\rmH \mathbf{A}) \mathbf{A} =
(|\Axl + \Axp|^2 + |\Ayl + \Ayp|^2)
\begin{pmatrix}
\Axl + \Axp \\ \Ayl + \Ayp
\end{pmatrix}.
\end{align}
Our intention is to study $\Ax$ and we will consider the two cases
that either (i)~$|\Axl|$ and $|\Ayl|$ are of the same order of
magnitude (transmission using polarization multiplexing) or (ii)~$\Ayl
= 0$ (single-polarization transmission). The first assumption above
can then be strictly formulated as $|\Axp| \ll |\Axl|$ and $|\Ayp| \ll
|\Axl|$.  This allows us to approximate the nonlinear term to leading
order according to
\begin{align}
(\mathbf{A}^\rmH \mathbf{A}) \mathbf{A} \approx
(|\Axl|^2 + |\Ayl|^2)
\begin{pmatrix}
\Axl \\ \Ayl
\end{pmatrix}.
\end{align}
Inserting this into the Manakov equation, we obtain
\begin{align}
i \pd{}{z}
\begin{pmatrix}
\Axl + \Axp \\ \Ayl + \Ayp
\end{pmatrix} = \frac{\beta_2}{2} \pdn{}{t}{2}
\begin{pmatrix}
\Axl + \Axp \\ \Ayl + \Ayp
\end{pmatrix} - \gamma (|\Axl|^2 + |\Ayl|^2)
\begin{pmatrix}
\Axl \\ \Ayl
\end{pmatrix} + i \frac{g - \alpha}{2}
\begin{pmatrix}
\Axl + \Axp \\ \Ayl + \Ayp
\end{pmatrix},
\end{align}
where the $z$-dependence is now implicit for compactness.  The fact
that $\mathbf{A}_l$ solves (\ref{eq_manakov}) when $\gamma = 0$,
implies that
\begin{align}
i \pd{}{z}
\begin{pmatrix}
\Axp \\ \Ayp
\end{pmatrix} = \frac{\beta_2}{2} \pdn{}{t}{2}
\begin{pmatrix}
\Axp \\ \Ayp
\end{pmatrix} - \gamma (|\Axl|^2 + |\Ayl|^2)
\begin{pmatrix}
\Axl \\ \Ayl
\end{pmatrix} + i \frac{g - \alpha}{2}
\begin{pmatrix}
\Axp \\ \Ayp
\end{pmatrix}.
\end{align}
%It is seen that the linearly propagating field gives rise to
%perturbations in both the $x$ and $y$ polarizations, but in this
%approximate model, the interaction of these perturbations is
%neglected.
%
We will study the $x$-polarized perturbation, which is described by
\opt{\begin{align} i \pd{\Axp}{z} = \frac{\beta_2}{2}
\pdn{\!\Axp}{t}{2}
- \gamma (|\Axl|^2 + |\Ayl|^2) \Axl + i \frac{g - \alpha}{2} \Axp.
\end{align}
or}
\begin{align}
\pd{\Axp}{z} = -i \frac{\beta_2}{2} \pdn{\!\Axp}{t}{2}
+ i \gamma (|\Axl|^2 + |\Ayl|^2) \Axl + \frac{g - \alpha}{2} \Axp.
\end{align}
For compactness, we temporarily introduce $S(z, t) = i \gamma
(|\Axl|^2 + |\Ayl|^2) \Axl$ to denote the source term in the partial
differential equation for the perturbation, which is then written
\begin{align}
\pd{\Axp}{z} + i \frac{\beta_2}{2} \pdn{\!\Axp}{t}{2} - \frac{g -
\alpha}{2} \Axp = S.
\end{align}

\subsection{Formal solution}
\label{sec_formal_solution}
We now derive a general formal solution without any assumptions about
the input signal or the system parameters. To describe the power
evolution, we introduce $P(z)$ as a function that satisfies the
equation
\begin{align}
\label{eq_alpha_def}
\od{P}{z} = [g(z) - \alpha(z)] P.
\end{align}
In the absence of any Raman amplification this function varies as
$e^{-\alpha z}$ between the amplifiers.  The EDFA gain can be modeled
by a $\delta$-function in $g(z)$ to obtain the discontinuities in
$P(z)$ at each $z$ corresponding to the location of an amplifier.
Introducing $\Axp(z, t) = \sqrt{P(z)} \, \psi(z, t)$, we obtain the
equation
\begin{align}
\frac{g - \alpha}{2} \sqrt{P} \psi + \sqrt{P} \pd{\psi}{z} + i
\frac{\beta_2}{2} \sqrt{P} \pdn{\psi}{t}{2} - \frac{g - \alpha}{2}
\sqrt{P} \psi = S,
\end{align}
or
\begin{align}
\pd{\psi}{z} + i \frac{\beta_2}{2} \pdn{\psi}{t}{2} =
\frac{S}{\sqrt{P}}.
\end{align}
We proceed by describing the accumulated dispersion, $B(z)$, by a
function that satisfies
\begin{align}
\od{B}{z} = \beta_2(z).
\end{align}
If there is lumped dispersion compensation (such as a chirped fiber
Bragg grating), then $B$ will have discontinuities.  This can be
modeled by a $\delta$-function in $\beta_2(z)$ to obtain
\begin{align}
B(z) = \int_{0}^{z} \beta_2(\zeta) \, d\zeta.
\end{align}
Fourier transforming\footnote{We use the same definition for the
  Fourier transform as Proakis~\cite{proakis_dig_comm_5ed}, i.e.,
  $\tilde{u}(f) = \int_{-\infty}^{\infty} u(t) e^{-i 2 \pi f t} dt$
  and $u(t) = \int_{-\infty}^{\infty} \tilde{u}(f) e^{i 2 \pi f t}
  df$.} the equation, and denoting this operation by tilde, we obtain
\begin{align}
\pd{\tilde{\psi}}{z} - i (2 \pi f)^2 \frac{\beta_2}{2} \tilde{\psi} =
\frac{\tilde{S}}{\sqrt{P}},
\end{align}
which can be written
\begin{align}
\pd{}{z} \left( e^{-i (2 \pi f)^2 B/2} \tilde{\psi} \right) = e^{-i (2
  \pi f)^2 B/2} \frac{\tilde{S}}{\sqrt{P}}.
\end{align}
Integrating $\int_{0}^{z} \cdot \, d\zeta$, we use
\begin{align}
\left[ e^{-i (2 \pi f)^2 B/2} \tilde{\psi} \right]_{\zeta = 0}^{\zeta
  = z} = \{ \tilde{\psi}(0, f) = 0 \} = e^{-i (2 \pi f)^2 B(z)/2}
  \tilde{\psi} (z, f)
\end{align}
to obtain
\begin{align}
\tilde{\psi} (z, f) = e^{i (2 \pi f)^2 B(z)/2} \int_{0}^{z} e^{-i (2
  \pi f)^2 B(\zeta)/2} \frac{\tilde{S}(\zeta, f)}{\sqrt{P(\zeta)}} \,
  d\zeta.
\end{align}
The final perturbation at the receiver is described by $\tilde{\psi}
(\Ltot, f)$, where $\Ltot$ is the total system length (possibly
containing many spans with fibers and amplifiers).  Assuming that
perfect EDC is performed, the exponential before the integration is
canceled and we obtain
\begin{align}
\tilde{\psi} (\Ltot, f) = \int_{0}^{\Ltot} e^{-i (2 \pi f)^2 B(z)/2}
\frac{\tilde{S}(z, f)}{\sqrt{P(z)}} \, dz.
\end{align}
Assuming that the power at $z = \Ltot$ is equal to the power at the
transmitter, $P_0$, we have $\Axp(\Ltot, t) = \sqrt{P(\Ltot)} \,
\psi(\Ltot, t) = \sqrt{P_0} \, \psi(\Ltot, t)$ and we can write the
general expression for the perturbation as
\begin{align}
\Axptilde (\Ltot, f) = \int_{0}^{\Ltot} e^{-i (2 \pi f)^2 B(z)/2}
\frac{\tilde{S}(z, f)}{\sqrt{P(z)}} \sqrt{P_0} \, dz =
\int_{0}^{\Ltot} e^{-i (2 \pi f)^2 B(z)/2} \frac{\tilde{S}(z,
f)}{\sqrt{p(z)}} \, dz,
\end{align}
where $p(z) = P(z)/P_0$ has been introduced.  This function describes
the normalized power evolution through the system.  Using the
definition of $S$, we find
\begin{align}
\label{eq_pert_sol}
\Axptilde (\Ltot, f) = \int_{0}^{\Ltot} \frac{e^{-i (2 \pi f)^2
B(z)/2}}{\sqrt{p(z)}} \fourier[i \gamma (|\Axl|^2 + |\Ayl|^2) \Axl] \,
dz,
\end{align}
where $\fourier$ denotes Fourier transformation.

\section{Signal model}
\label{section_signal_model}
Provided that the first assumption is fulfilled, (\ref{eq_pert_sol})
is the solution to (\ref{eq_manakov}) without further approximations.
However, we need to choose the system parameters, select the boundary
condition at $z = 0$, i.e., the input signal, and perform the
integration.  When trying to do this with an accurate modeling of the
signal pulses, the calculations become cumbersome and the final
expression needs to be averaged over the data.  We here instead use
the second assumption, i.e., we write the initial field in the way
suggested in~\cite{poggiolini_2011_ptl2}.  This model is
\begin{align}
\label{eq_signal_model_x}
\Axltilde(0, f) &= \sqrt{f_0}
\sum_{\kx = -\infty}^{\infty} \xikx \sqrt{\Gx(\kx f_0)} \, \delta(f -
\kx f_0),\\
\label{eq_signal_model_y}
\Ayltilde(0, f) &= \sqrt{f_0}
\sum_{\ky = -\infty}^{\infty} \xiky \sqrt{\Gy(\ky f_0)} \, \delta(f -
\ky f_0),
\end{align}
where $\xi_k$ and $\zeta_k$ are complex independent Gaussian random
variables of unit variance and the input signal PSD of the $x$ and $y$
polarizations are denoted by $\Gx(f)$ and $\Gy(f)$, respectively.  In
the following, we will suppress the infinite summation limits for
notational convenience.  We account for dispersion and power
variations during the propagation by modifying this to
\begin{align}
\Axltilde(z, f) &= \sqrt{f_0 p(z)}
\sum_{\kx}^{} \xikx \sqrt{\Gx(\kx f_0)} \, \delta(f - \kx f_0) e^{i (2
\pi \kx f_0)^2 B(z)/2},\\
\Ayltilde(z, f) &= \sqrt{f_0 p(z)}
\sum_{\ky}^{} \xiky \sqrt{\Gy(\ky f_0)} \, \delta(f - \ky f_0) e^{i (2
\pi \ky f_0)^2 B(z)/2}.
\end{align}
Using this linear solution, we can now calculate the perturbation
using (\ref{eq_pert_sol}).  In order to do this we need the expression
\begin{align}
\fourier[i \gamma (|\Axl|^2 + |\Ayl|^2) \Axl] = i \gamma
\fourier[\Axl^2 \Axl^*] + i \gamma \fourier[\Axl \Ayl \Ayl^*].
\end{align}
The solution (\ref{eq_pert_sol}) can be written
\begin{align}
\label{eq_pert_sol_two_ints}
\Axptilde (\Ltot, f) = i \int_{0}^{\Ltot} \gamma e^{-i (2 \pi f)^2
  B/2} \frac{\fourier[\Axl^2 \Axl^*]}{\sqrt{p}} \, dz + i
  \int_{0}^{\Ltot} \gamma e^{-i (2 \pi f)^2 B/2} \frac{\fourier[\Axl
  \Ayl \Ayl^*]}{\sqrt{p}} \, dz,
\end{align}
where we again omit the $z$ dependence for notational compactness.  We
have
\begin{align}
\Axl(z, t) &= \sqrt{f_0 p}
\sum_{\kx}^{} \xikx \sqrt{\Gx(\kx f_0)} \, e^{i (2 \pi \kx f_0) t}
e^{i (2 \pi \kx f_0)^2 B/2},\\
\Ayl(z, t) &= \sqrt{f_0 p}
\sum_{\ky}^{} \xiky \sqrt{\Gy(\ky f_0)} \, e^{i (2 \pi \ky f_0) t}
e^{i (2 \pi \ky f_0)^2 B/2},
\end{align}
and get
\begin{align}
\Axl^2 \Axl^* &= \left( \sqrt{f_0 p}
\sum_{\kx}^{} \xikx \sqrt{\Gx(\kx f_0)} \, e^{i (2 \pi \kx f_0) t}
e^{i (2 \pi \kx f_0)^2 B/2} \right) \left( \sqrt{f_0 p}
\sum_{\lx}^{} \xilx \sqrt{\Gx(\lx f_0)} \, e^{i (2 \pi \lx f_0) t}
e^{i (2 \pi \lx f_0)^2 B/2} \right) \nonumber \\
& \quad \times \left( \sqrt{f_0 p}
\sum_{\mx}^{} \ximx^* \sqrt{\Gx(\mx f_0)} \, e^{-i (2 \pi \mx f_0) t}
e^{-i (2 \pi \mx f_0)^2 B/2} \right), \\
\Axl \Ayl \Ayl^* &= \left( \sqrt{f_0 p}
\sum_{\kx}^{} \xikx \sqrt{\Gx(\kx f_0)} \, e^{i (2 \pi \kx f_0) t}
  e^{i (2 \pi \kx f_0)^2 B/2} \right) \left( \sqrt{f_0 p}
\sum_{\ly}^{} \xily \sqrt{\Gy(\ly f_0)} \, e^{i (2 \pi \ly f_0) t}
e^{i (2 \pi \ly f_0)^2 B/2} \right) \nonumber \\
& \quad \times \left( \sqrt{f_0 p}
\sum_{\my}^{} \ximy^* \sqrt{\Gy(\my f_0)} \, e^{-i (2 \pi \my f_0) t}
e^{-i (2 \pi \my f_0)^2 B/2} \right).
\end{align}
These expressions can be simplified to
\begin{align}
\Axl^2 \Axl^* &= (f_0 p)^{3/2}
\sum_{\kx, \lx, \mx}^{} \xikx \xilx \ximx^* \sqrt{\Gx(\kx f_0) \Gx(\lx
f_0) \Gx(\mx f_0)} e^{i 2 \pi (\kx + \lx - \mx) f_0 t} e^{i (2 \pi)^2
(\kx^2 + \lx^2 - \mx^2) f_0^2 B/2}, \\
\Axl \Ayl \Ayl^* &= (f_0 p)^{3/2}
\sum_{\kx, \lx, \mx}^{} \xikx \xily \ximy^* \sqrt{\Gx(\kx f_0) \Gy(\lx
f_0) \Gy(\mx f_0)} e^{i 2 \pi (\kx + \lx - \mx) f_0 t} e^{i (2 \pi)^2
(\kx^2 + \lx^2 - \mx^2) f_0^2 B/2}.
\end{align}
The Fourier transforms are
\begin{align}
\fourier[\Axl^2 \Axl^*] &= (f_0 p)^{3/2}
\sum_{\kx, \lx, \mx}^{}
\xikx \xilx \ximx^* \sqrt{\Gx(\kx f_0) \Gx(\lx f_0) \Gx(\mx f_0)}
\delta(f - (\kx + \lx - \mx) f_0) e^{i (2 \pi)^2 (\kx^2 + \lx^2 -
\mx^2) f_0^2 B/2},\\
\!\!\fourier[\Axl \Ayl \Ayl^*] &= (f_0 p)^{3/2}
\sum_{\kx, \lx, \mx}^{} \xikx \xily \ximy^* \sqrt{\Gx(\kx f_0) \Gy(\lx
  f_0) \Gy(\mx f_0)} \delta(f - (\kx + \lx - \mx) f_0) e^{i (2 \pi)^2
  (\kx^2 + \lx^2 - \mx^2) f_0^2 B/2}.
\end{align}
The integrands of (\ref{eq_pert_sol_two_ints}) can then be written
\begin{align}
\gamma e^{-i (2 \pi f)^2 B/2} \frac{\fourier[\Axl^2 \Axl^*]}{\sqrt{p}}
&=
\gamma e^{-i (2 \pi f)^2 B/2} f_0^{3/2} p
\sum_{\kx, \lx, \mx}^{} \xikx \xilx \ximx^* \sqrt{\Gx(\kx f_0) \Gx(\lx
f_0) \Gx(\mx f_0)} \nonumber \\
& \quad \times \delta(f - (\kx + \lx - \mx) f_0) e^{i (2 \pi)^2 (\kx^2
+ \lx^2 - \mx^2) f_0^2 B/2}, \\
\gamma e^{-i (2 \pi f)^2 B/2} \frac{\fourier[\Axl \Ayl
\Ayl^*]}{\sqrt{p}} &=
\gamma e^{-i (2 \pi f)^2 B/2} f_0^{3/2} p
\sum_{\kx, \lx, \mx}^{} \xikx \xily \ximy^* \sqrt{\Gx(\kx f_0) \Gy(\lx
f_0) \Gy(\mx f_0)} \nonumber \\
& \quad \times \delta(f - (\kx + \lx - \mx) f_0) e^{i (2 \pi)^2 (\kx^2
+ \lx^2 - \mx^2) f_0^2 B/2}.
\end{align}
We can move the exponential functions containing $B$ after the
summation signs to obtain
\begin{align}
\gamma e^{-i (2 \pi f)^2 B/2} \frac{\fourier[\Axl^2 \Axl^*]}{\sqrt{p}}
&=
f_0^{3/2} \gamma p
\sum_{\kx, \lx, \mx}^{} \xikx \xilx \ximx^* \sqrt{\Gx(\kx f_0) \Gx(\lx
f_0) \Gx(\mx f_0)} \nonumber \\
& \quad \times \delta(f - (\kx + \lx - \mx) f_0) e^{i (2 \pi)^2 (\kx^2
+ \lx^2 - \mx^2) f_0^2 B/2} e^{-i (2 \pi)^2 (\kx + \lx - \mx)^2 f_0^2
B/2}, \\
\gamma e^{-i (2 \pi f)^2 B/2} \frac{\fourier[\Axl \Ayl \Ayl^*]}{\sqrt{p}}
&=
f_0^{3/2} \gamma p
\sum_{\kx, \lx, \mx}^{} \xikx \xily \ximy^* \sqrt{\Gx(\kx f_0) \Gy(\lx
f_0) \Gy(\mx f_0)} \nonumber \\
& \quad \times \delta(f - (\kx + \lx - \mx) f_0) e^{i (2 \pi)^2 (\kx^2
+ \lx^2 - \mx^2) f_0^2 B/2} e^{-i (2 \pi)^2 (\kx + \lx - \mx)^2 f_0^2
B/2}.
\end{align}
We use that
\begin{align}
(\kx^2 + \lx^2 - \mx^2) - (\kx + \lx - \mx)^2 = -2 (\kx - \mx) (\lx -
  \mx)
\end{align}
to obtain
\begin{align}
\gamma e^{-i (2 \pi f)^2 B/2} \frac{\fourier[\Axl^2 \Axl^*]}{\sqrt{p}} &=
f_0^{3/2} \gamma p
\sum_{\kx, \lx, \mx}^{} \xikx \xilx \ximx^* \sqrt{\Gx(\kx f_0) \Gx(\lx
f_0) \Gx(\mx f_0)} \nonumber \\
& \quad \times \delta(f - (\kx + \lx - \mx) f_0)e^{-i 4 \pi^2 (\kx -
\mx) (\lx - \mx) f_0^2 B} \\
\gamma e^{-i (2 \pi f)^2 B/2} \frac{\fourier[\Axl \Ayl \Ayl^*]}{\sqrt{p}}
&=
f_0^{3/2} \gamma p
\sum_{\kx, \lx, \mx}^{} \xikx \xily \ximy^* \sqrt{\Gx(\kx f_0) \Gy(\lx
f_0) \Gy(\mx f_0)} \nonumber \\
& \quad \times \delta(f - (\kx + \lx - \mx) f_0) e^{-i 4 \pi^2 (\kx -
\mx) (\lx - \mx) f_0^2 B}.
\end{align}
The solution can therefore be written
\begin{align}
\Axptilde (\Ltot, f) &=
i \int_{0}^{\Ltot} f_0^{3/2} \gamma p
\sum_{\kx, \lx, \mx}^{} \xikx \xilx \ximx^* \sqrt{\Gx(\kx f_0) \Gx(\lx
f_0) \Gx(\mx f_0)} \, \delta(f - (\kx + \lx - \mx) f_0) e^{-i 4 \pi^2
(\kx - \mx) (\lx - \mx) f_0^2 B} \, dz \nonumber \\
& \quad + i \int_{0}^{\Ltot} f_0^{3/2} \gamma p
\sum_{\kx, \lx, \mx}^{} \xikx \xily \ximy^* \sqrt{\Gx(\kx f_0) \Gy(\lx
f_0) \Gy(\mx f_0)} \, \delta(f - (\kx + \lx - \mx) f_0) e^{-i 4 \pi^2
(\kx - \mx) (\lx - \mx) f_0^2 B} \, dz.
\end{align}
Inverse Fourier transformation gives
\begin{align}
\Axp (\Ltot, t) &=
i f_0^{3/2} \int_{0}^{\Ltot} \gamma p
\sum_{\kx, \lx, \mx}^{} \xikx \xilx \ximx^* \sqrt{\Gx(\kx f_0) \Gx(\lx
f_0) \Gx(\mx f_0)}
e^{i 2 \pi (\kx + \lx - \mx) f_0 t} e^{-i 4 \pi^2 (\kx - \mx) (\lx -
  \mx) f_0^2 B} \, dz \nonumber \\
& \quad + i f_0^{3/2} \int_{0}^{\Ltot} \gamma p
\sum_{\kx, \lx, \mx}^{} \xikx \xily \ximy^* \sqrt{\Gx(\kx f_0) \Gy(\lx
f_0) \Gy(\mx f_0)}
e^{i 2 \pi (\kx + \lx - \mx) f_0 t} e^{-i 4 \pi^2 (\kx - \mx) (\lx -
  \mx) f_0^2 B} \, dz.
\end{align}
We can also rearrange integration and summation to obtain
\begin{align}
\Axp (\Ltot, t) &=
i f_0^{3/2}
\sum_{\kx, \lx, \mx} \xikx \xilx \ximx^* \sqrt{\Gx(\kx f_0) \Gx(\lx
f_0) \Gx(\mx f_0)} e^{i 2 \pi (\kx + \lx - \mx) f_0 t}
\int_{0}^{\Ltot} \gamma p e^{-i 4 \pi^2 (\kx - \mx) (\lx - \mx) f_0^2
B} \, dz\nonumber \\
& \quad + i f_0^{3/2}
\sum_{\kx, \lx, \mx} \xikx \xily \ximy^* \sqrt{\Gx(\kx f_0) \Gy(\lx
f_0) \Gy(\mx f_0)} e^{i 2 \pi (\kx + \lx - \mx) f_0 t}
\int_{0}^{\Ltot} \gamma p e^{-i 4 \pi^2 (\kx - \mx) (\lx - \mx) f_0^2
B} \, dz.
\end{align}
Considering $\gamma(z)$, $p(z)$, $B(z)$, and $\Ltot$ to be given
system parameters, we introduce
\begin{align}
\Cklm \equiv \mathcal{C}(k f_0, l f_0, m f_0) \equiv \int_{0}^{\Ltot}
\gamma p e^{-i 4 \pi^2 (\kx - \mx) (\lx - \mx) f_0^2 B} \, dz
\end{align}
to write this as
\begin{align}
\label{eq_pert_with_signal_model}
\Axp (\Ltot, t) &=
i f_0^{3/2}
\sum_{\kx, \lx, \mx} \Cklm \xikx \xilx \ximx^* \sqrt{\Gx(\kx f_0)
\Gx(\lx f_0) \Gx(\mx f_0)} e^{i 2 \pi (k + l - m) f_0 t} \nonumber \\
& \quad + i f_0^{3/2}
\sum_{\kx, \lx, \mx} \Cklm \xikx \xily \ximy^* \sqrt{\Gx(\kx f_0)
\Gy(\lx f_0) \Gy(\mx f_0)} e^{i 2 \pi (k + l - m) f_0 t} \nonumber \\
&= i f_0^{3/2}
\sum_{\kx, \lx, \mx} \Cklm e^{i 2 \pi (k + l - m) f_0 t} \xikx
  \sqrt{\Gx(\kx f_0)} \left( \xilx \ximx^* \sqrt{\Gx(\lx f_0) \Gx(\mx
  f_0)} + \xily \ximy^* \sqrt{\Gy(\lx f_0) \Gy(\mx f_0)}\right).
\end{align}
For later use we notice that
\begin{align}
|\Cklm| \le \int_{0}^{\Ltot} \left| \gamma p e^{-i 4 \pi^2 (\kx - \mx)
  (\lx - \mx) f_0^2 B} \right| \, dz =
\int_{0}^{\Ltot} \gamma p \, dz \le
\hat{\gamma} \hat{p} L,
\end{align}
where $\hat{p}$ and $\hat{\gamma}$ are the maximum values of $p(z)$
and $\gamma(z)$ in the interval $z \in [0, L]$.  This means that
$|\Cklm|$ is upper bounded by a constant as $k$, $l$, $m$, and $f_0$
are changed.

\section{Power spectral density}
\label{section_psd}
When the input signal is modeled by (\ref{eq_signal_model_x}) and
(\ref{eq_signal_model_y}), then the perturbation is given by
(\ref{eq_pert_with_signal_model}).  The next step is to calculate the
corresponding PSD.  Using the Wiener-Khinchin
theorem~\cite[p.~67]{proakis_dig_comm_5ed}, we have that the PSD is
the Fourier transform of the autocorrelation according to
\begin{align}
\Gxptilde(f) = \fourier[R(\tau)] = \infint R(\tau) e^{-i 2 \pi f \tau}
\, d\tau,
\end{align}
where
\begin{align}
R = \expval\{\Axp(L, t_1) \Axp^*(L, t_2)\}
\end{align}
and $\expval\{\cdot\}$ denotes the expectation operator.  As will be
shown, the latter expression can be written as a function of the time
difference $\tau = t_1 - t_2$. We will suppress the infinite limits
for notational convenience.  We have
\begin{align}
&\Axp(L, t_1) \Axp^*(L, t_2) = \\
&\quad \left[ i f_0^{3/2}
\sum_{\kx, \lx, \mx}^{}
\Cklm e^{i 2 \pi (\kx + \lx - \mx) f_0 t_1} \xikx \sqrt{\Gx(\kx f_0)}
\left( \xilx \ximx^* \sqrt{\Gx(\lx f_0) \Gx(\mx f_0)} + \xily \ximy^*
\sqrt{\Gy(\lx f_0) \Gy(\mx f_0)}\right) \right] \times \nonumber \\
& \quad \left[ -i f_0^{3/2}
\sum_{\kxx, \lxx, \mxx}^{} \CCklm^* e^{-i 2 \pi (\kxx + \lxx - \mxx)
      f_0 t_2} \xikxx^* \sqrt{\Gx(\kxx f_0)} \left( \xilxx^* \ximxx
      \sqrt{\Gx(\lxx f_0) \Gx(\mxx f_0)} + \xilyy^* \ximyy
      \sqrt{\Gy(\lxx f_0) \Gy(\mxx f_0)}\right) \right] = \nonumber \\
& \quad f_0^3
\sum_{\substack{\kx,\lx,\mx\\\kxx,\lxx,\mxx}} \Cklm \CCklm^* e^{i 2
    \pi (\kx + \lx - \mx) f_0 t_1} e^{-i 2 \pi (\kxx + \lxx - \mxx)
    f_0 t_2} \xikx \xikxx^* \sqrt{\Gx(\kx f_0)} \sqrt{\Gx(\kxx f_0)}
    \times \nonumber \\
& \quad \left( \xilx \ximx^* \sqrt{\Gx(\lx f_0) \Gx(\mx f_0)} + \xily
  \ximy^* \sqrt{\Gy(\lx f_0) \Gy(\mx f_0)}\right) \left( \xilxx^*
  \ximxx \sqrt{\Gx(\lxx f_0) \Gx(\mxx f_0)} + \xilyy^* \ximyy
  \sqrt{\Gy(\lxx f_0) \Gy(\mxx f_0)}\right). \nonumber
\end{align}
We see that the complete autocorrelation function will consist of four
terms
\begin{align}
R = \Ra + \Rb + \Rc + \Rd
\end{align}
where
\begin{align}
\label{eq_r1}
\Ra &= \expval \Bigg\{ f_0^3
\sum_{\substack{\kx,\lx,\mx\\\kxx,\lxx,\mxx}} \Cklm \CCklm^* e^{i 2
    \pi (\kx + \lx - \mx) f_0 t_1} e^{-i 2 \pi (\kxx + \lxx - \mxx)
    f_0 t_2} \nonumber \\
& \quad \times \xikx \xilx \ximx^* \xikxx^* \xilxx^* \ximxx
    \sqrt{\Gx(\kx f_0) \Gx(\lx f_0) \Gx(\mx f_0) \Gx(\kxx f_0)
    \Gx(\lxx f_0) \Gx(\mxx f_0)} \Bigg\},\\
\Rb &= \expval \Bigg\{ f_0^3
\sum_{\substack{\kx,\lx,\mx\\\kxx,\lxx,\mxx}} \Cklm \CCklm^* e^{i 2
    \pi (\kx + \lx - \mx) f_0 t_1} e^{-i 2 \pi (\kxx + \lxx - \mxx)
    f_0 t_2} \nonumber \\
& \quad \times \xikx \xilx \ximx^* \xikxx^* \xilyy^* \ximyy
    \sqrt{\Gx(\kx f_0) \Gx(\lx f_0) \Gx(\mx f_0) \Gx(\kxx f_0)
    \Gy(\lxx f_0) \Gy(\mxx f_0)} \Bigg\},\\
\Rc &= \expval \Bigg\{ f_0^3
\sum_{\substack{\kx,\lx,\mx\\\kxx,\lxx,\mxx}} \Cklm \CCklm^* e^{i 2
    \pi (\kx + \lx - \mx) f_0 t_1} e^{-i 2 \pi (\kxx + \lxx - \mxx)
    f_0 t_2} \nonumber \\
& \quad \times \xikx \xily \ximy^* \xikxx^* \xilxx^* \ximxx
    \sqrt{\Gx(\kx f_0) \Gy(\lx f_0) \Gy(\mx f_0) \Gx(\kxx f_0)
    \Gx(\lxx f_0) \Gx(\mxx f_0)} \Bigg\},\\
\label{eq_r4}
\Rd &= \expval \Bigg\{ f_0^3
\sum_{\substack{\kx,\lx,\mx\\\kxx,\lxx,\mxx}} \Cklm \CCklm^* e^{i 2
    \pi (\kx + \lx - \mx) f_0 t_1} e^{-i 2 \pi (\kxx + \lxx - \mxx)
    f_0 t_2} \nonumber \\
& \quad \times \xikx \xily \ximy^* \xikxx^* \xilyy^* \ximyy
    \sqrt{\Gx(\kx f_0) \Gy(\lx f_0) \Gy(\mx f_0) \Gx(\kxx f_0)
    \Gy(\lxx f_0) \Gy(\mxx f_0)} \Bigg\}.
\end{align}
These four terms in the autocorrelation will give rise to four
different terms in the total PSD, which we denote by $\Gxptildea$,
$\Gxptildeb$, $\Gxptildec$, and $\Gxptilded$, respectively.  We need
to find these four terms individually.

\subsection{The random variables}
It is seen that in (\ref{eq_r1})--(\ref{eq_r4}), all terms except the
$\xi$ and $\zeta$ are deterministic.  In order to carry out the
expectation operation, in this section we therefore need to
investigate these random variables.

The $\xi$ and $\zeta$ are complex independent Gaussian random
variables of unit variance, i.e., $\expval\{\xikx\} = 0$,
$\expval\{\xikx^2\} = 0$, $\expval\{|\xikx|^2\} = 1$, $\forall k$.
Analogous expressions hold for $\xiky$. Using this, we can simplify
the six-dimensional sums (\ref{eq_r1})--(\ref{eq_r4}) in the following
way.  First assume that one of the summation variables, say $\kx$, has
a value different from all other summation variables, i.e., $\kx$ is
unique.  Due to the independence, we then have
\begin{align}
\expval\{\xikx \xilx \ximx^* \xikxx^* \xilxx^* \ximxx\} =
\expval\{\xikx\} \expval\{\xilx \ximx^* \xikxx^* \xilxx^* \ximxx\} =
0.
\end{align}
An identical argument holds if some of the $\xi$ are replaced by
$\zeta$.  This implies that the expected value is always zero when one
of the summation variables is unique.  Thus, we have the condition
that no summation variable can be unique.  Second we assume that no
summation variable is unique, but there are three pairwise equal
values, where each pair has a unique value.  Assume for example that
$\kx = \lx$, $\kxx = \lxx$, $\mx = \mxx$.  Then
\begin{align}
\expval\{\xikx \xilx \ximx^* \xikxx^* \xilxx^* \ximxx\} &=
\expval\{\xikx \xikx \ximx^* \xikxx^* \xikxx^* \ximx\} =
\expval\{\xikx \xikx\} \expval\{\xikxx^* \xikxx^*\} \expval\{\ximx^*
\ximx\} =
\expval\{\xikx^2\} \expval\{(\xikxx^*)^2\} \expval\{|\ximx|^2\} = 0.
\end{align}
We conclude that each random variable must be paired up with the
complex conjugated version of the same random variable.  This
conclusion reduces the dimensionality of the summation to three or
less. However, the dimensionality cannot be less than three, because
then the expression $\Gxptilde \to 0$ as $f_0 \to 0$.  To see this we
first notice that
\begin{align}
|\Gxptilde| \le |\Gxptildea| + |\Gxptildeb| + |\Gxptildec| +
|\Gxptilded|,
\end{align}
and since all terms behave similarly in this respect, we can study,
say, $\Gxptildea$. Let us select the one-dimensional case $\kx = \lx =
\mx = \kxx = \lxx = \mxx$. We then have
\begin{align}
\Ra \opt{&= \expval \Bigg\{ f_0^3
\sum_{\substack{\kx,\lx,\mx\\\kxx,\lxx,\mxx}} \Cklm \CCklm^* e^{i 2
    \pi (\kx + \lx - \mx) f_0 t_1} e^{-i 2 \pi (\kxx + \lxx - \mxx)
    f_0 t_2} \nonumber \\
& \quad \times \xikx \xilx \ximx^* \xikxx^* \xilxx^* \ximxx
    \sqrt{\Gx(\kx f_0) \Gx(\lx f_0) \Gx(\mx f_0) \Gx(\kxx f_0)
    \Gx(\lxx f_0) \Gx(\mxx f_0)} \Bigg\} \nonumber \\}
&= \expval \Bigg\{ f_0^3
\sum_{\kx} |\mathcal{C}_{kkk}|^2 e^{i 2 \pi \kx f_0 \tau} |\xikx|^6
    \Gx^3(\kx f_0) \Bigg\} = f_0^3
\sum_{\kx} |\mathcal{C}_{kkk}|^2 e^{i 2 \pi \kx f_0 \tau} \expval \{
  |\xikx|^6\} \Gx^3(\kx f_0).
\end{align}
We get
\begin{align}
\Gxptildea &= f_0^3
\sum_{\kx} |\mathcal{C}_{kkk}|^2 \expval \{ |\xikx|^6 \} \Gx^3(\kx
f_0) \delta(f - k f_0).
\end{align}
We see that $\Gxptildea \to 0$ as $f_0 \to 0$ by identifying the
Riemann sum and writing the expression as
\begin{align}
\Gxptildea &= f_0^2 \int |\mathcal{C}(f_1, f_1, f_1)|^2 \expval \{
|\xikx|^6 \} \Gx^3(f_1) \delta(f - f_1) \, df_1 = f_0^2
|\mathcal{C}(f, f, f)|^2 \expval \{ |\xikx|^6 \} \Gx^3(f).
\end{align}
Remembering that the value of $|\mathcal{C}|^2$ is upper bounded by a
constant, the fact that $\Gxptildea \to 0$ is now obvious.  An
analogous argument can be made for the case of a two-dimensional sum.
We conclude that the summation must have dimension exactly three,
i.e., each random variable must be paired up with the complex
conjugated version of the same random variable and all three pairs
must have unique values.

\subsection{The first term in the PSD}
We now study the first expression, i.e., $\Gxptildea(f) =
\fourier[\Ra(\tau)]$.  Following the above rules for how the indices
can be chosen, we have six possible combinations:
\begin{align}
\begin{array}{ccc}
\kx = \kxx & \lx = \lxx & \mx = \mxx,\\
\kx = \kxx & \lx = \mx & \mxx = \lxx,\\
\kx = \lxx & \lx = \kxx & \mx = \mxx,\\
\kx = \lxx & \lx = \mx & \mxx = \kxx,\\
\kx = \mx & \lx = \kxx & \mxx = \lxx,\\
\kx = \mx & \lx = \lxx & \mxx = \kxx.
\end{array}
\end{align}
We need to study these different possibilities individually.

\subsubsection{The case $\kx = \kxx$, $\lx = \lxx$, $\mx = \mxx$}
We then have
\begin{align}
\Ra &= \expval \left\{ f_0^3
\sum_{\kx,\lx,\mx} |\Cklm|^2 e^{i 2 \pi (\kx + \lx - \mx) f_0 t_1}
    e^{-i 2 \pi (\kx + \lx - \mx) f_0 t_2} |\xikx|^2 |\xilx|^2
    |\ximx|^2\Gx(\kx f_0) \Gx(\lx f_0) \Gx(\mx f_0) \right\} \nonumber
    \\
&= f_0^3
\sum_{\kx,\lx,\mx} |\Cklm|^2 e^{i 2 \pi (\kx + \lx - \mx) f_0 \tau}
  \Gx(\kx f_0) \Gx(\lx f_0) \Gx(\mx f_0),
\end{align}
\begin{align}
\Gxptildea(f) \opt{&= \int \Ra(\tau) e^{-i 2 \pi f \tau} \, d\tau
\nonumber \\
&= f_0^3
\sum_{\kx,\lx,\mx} |\Cklm|^2 \Gx(\kx f_0) \Gx(\lx f_0) \Gx(\mx f_0)
\int e^{i 2 \pi (\kx + \lx - \mx) f_0 \tau} e^{-i 2 \pi f \tau} \,
d\tau \nonumber \\
\opt{&= f_0^3
\sum_{\kx,\lx,\mx} |\Cklm|^2 \Gx(\kx f_0) \Gx(\lx f_0) \Gx(\mx f_0)
\int e^{-i 2 \pi [f - (\kx + \lx - \mx) f_0]\tau} \, d\tau
\nonumber \\}}
&= f_0^3
\sum_{\kx,\lx,\mx} |\Cklm|^2 \Gx(\kx f_0) \Gx(\lx f_0) \Gx(\mx f_0)
\delta(f - (\kx + \lx - \mx) f_0).
\end{align}
Letting $f_0 \to 0$, we get
\begin{align}
\Gxptildea(f) &= \iiint |\mathcal{C}(f_1, f_2, f_3)|^2 \Gx(f_1)
\Gx(f_2) \Gx(f_3) \delta(f - f_1 - f_2 + f_3)\, df_1 df_2 df_3
\nonumber \\
&= \iint |\mathcal{C}(f_1, f_2, f_1 + f_2 - f)|^2 \Gx(f_1) \Gx(f_2)
\Gx(f_1 + f_2 - f) \, df_1 df_2.
\end{align}

\subsubsection{The other cases}
Compared to the case above, the case $\kx = \lxx$, $\lx = \kxx$, $\mx
= \mxx$ is obtained by swapping $\kxx$ and $\lxx$. Since $\Cklm =
\Clkm$ and the rest of the expression is clearly invariant under this
change, this yields an identical result as the case above.  The other
four cases are all equivalent to the case $\kx = \kxx$, $\lx = \mx$,
$\mxx = \lxx$. We get
\begin{align}
\Ra &= \expval \left\{ f_0^3
\sum_{\kx\lx\lxx} \mathcal{C}_{\kx\lx\lx} \mathcal{C}_{\kx\lxx\lxx}^*
  e^{i 2 \pi (\kx + \lx - \lx) f_0 t_1} e^{-i 2 \pi (\kx + \lxx -
  \lxx) f_0 t_2} \xikx \xilx \xilx^* \xikx^* \xilxx^* \xilxx
  \sqrt{\Gx(\kx f_0) \Gx(\lx f_0) \Gx(\lx f_0) \Gx(\kx f_0) \Gx(\lxx
  f_0) \Gx(\lxx f_0)} \right\} \nonumber \\
&= \expval \left\{ f_0^3
\sum_{\kx\lx\lxx} \mathcal{C}_{\kx\lx\lx} \mathcal{C}_{\kx\lxx\lxx}^*
  e^{i 2 \pi \kx f_0 t_1} e^{-i 2 \pi \kx f_0 t_2} |\xikx|^2 |\xilx|^2
  |\xilxx|^2 \Gx(\kx f_0) \Gx(\lx f_0) \Gx(\lxx f_0) \right\} \nonumber
  \\
&= f_0^3
\sum_{\kx\lx\lxx} \mathcal{C}_{\kx\lx\lx} \mathcal{C}_{\kx\lxx\lxx}^*
e^{i 2 \pi \kx f_0 \tau} \Gx(\kx f_0) \Gx(\lx f_0) \Gx(\lxx f_0),
\end{align}
\begin{align}
\Gxptildea(f) \opt{&= \int \Ra(\tau) e^{-i 2 \pi f \tau} \, d\tau
\nonumber \\
&= f_0^3
\sum_{\kx\lx\lxx} \mathcal{C}_{\kx\lx\lx} \mathcal{C}^*_{\kx\lxx\lxx}
\Gx(\kx f_0) \Gx(\lx f_0) \Gx(\lxx f_0) \int e^{i 2 \pi \kx f_0
\tau} e^{-i 2 \pi f \tau} \, d\tau \nonumber \\}
&= f_0^3
\sum_{\kx\lx\lxx} \mathcal{C}_{\kx\lx\lx} \mathcal{C}^*_{\kx\lxx\lxx}
\Gx(\kx f_0) \Gx(\lx f_0) \Gx(\lxx f_0) \delta(f - \kx f_0).
\end{align}
Letting $f_0 \to 0$, we get
\begin{align}
\Gxptildea(f) &= \iiint \mathcal{C}(f_1, f_2, f_2) \mathcal{C}^*(f_1,
f_3, f_3) \Gx(f_1) \Gx(f_2) \Gx(f_3) \delta(f - f_1) \, df_1 df_2 df_3
\nonumber \\
&= \iint \mathcal{C}(f, f_2, f_2) \mathcal{C}^*(f, f_3, f_3) \Gx(f)
\Gx(f_2) \Gx(f_3) \, df_2 df_3.
\end{align}
However
\begin{align}
\mathcal{C}(f_1, f_2, f_2) = \mathcal{C}(f_1, f_3, f_3) =
\mathcal{C}(f_2, f_1, f_2) = \mathcal{C}(f_3, f_1, f_3) =
\int_{0}^{\Ltot} \gamma(z) p(z) \, dz \equiv \Cooo,
\end{align}
which is a real number. This gives
\begin{align}
\Gxptildea(f) &= \Cooo^2 \Gx(f) \iint \Gx(f_2) \Gx(f_3) \, df_2 df_3 =
\Cooo^2 \Gx(f) \int \Gx(f_2) \, df_2 \int \Gx(f_3) \, df_3 = \Cooo^2
P_x^2 \Gx(f).
\end{align}

\subsubsection{The total PSD for the first term}
The six possible index selection cases split into two groups of
degeneracy two and four, respectively. We get the total expression
\begin{align}
\Gxptildea(f) &= 2 \iint |\mathcal{C}(f_1, f_2, f_1 + f_2 - f)|^2
\Gx(f_1) \Gx(f_2) \Gx(f_1 + f_2 - f) \, df_1 df_2 + 4 \Cooo^2 P_x^2
\Gx(f).
\end{align}

\subsection{The second term in the PSD}
We now study the second expression, i.e., $\Gxptildeb(f) =
\fourier[\Rb(\tau)]$.  Following the above rules for how the indices
can be chosen, we have two combinations
\begin{align}
\begin{array}{ccc}
\kx = \kxx & \lx = \mx & \mxx = \lxx\\ \kx = \mx & \lx = \kxx & \mxx =
\lxx
\end{array}
\end{align}
The reason that we have fewer possibilities is that the expression
contains both $\xi$ and $\zeta$, which are independent.

\subsubsection{The case $\kx = \kxx$, $\lx = \mx$, $\mxx = \lxx$}
We then have
\begin{align}
\Rb &= \expval \left\{ f_0^3
\sum_{\kx,\lx,\lxx} \mathcal{C}_{\kx\lx\lx}
  \mathcal{C}_{\kx\lxx\lxx}^* e^{i 2 \pi \kx f_0 t_1} e^{-i 2 \pi \kxx
  f_0 t_2} |\xikx|^2 |\xilx|^2 |\xilyy|^2 \Gx(\kx f_0) \Gx(\lx f_0)
  \Gy(\lxx f_0) \right\} \nonumber \\
&= f_0^3
\sum_{\kx,\lx,\lxx} \mathcal{C}_{\kx\lx\lx}
\mathcal{C}_{\kx\lxx\lxx}^* e^{i 2 \pi \kx f_0 \tau} \Gx(\kx f_0)
\Gx(\lx f_0) \Gy(\lxx f_0),
\end{align}
\begin{align}
\Gxptildeb(f) &= f_0^3
\sum_{\kx,\lx,\lxx} \mathcal{C}_{\kx\lx\lx}
\mathcal{C}_{\kx\lxx\lxx}^* \Gx(\kx f_0) \Gx(\lx f_0) \Gy(\lxx f_0)
\delta(f - \kx f_0).
\end{align}
Letting $f_0 \to 0$, we get
\begin{align}
\Gxptildeb(f) &= \iiint \mathcal{C}(f_1, f_2, f_2) \mathcal{C}^*(f_1,
f_3, f_3) \Gx(f_1) \Gx(f_2) \Gy(f_3) \delta(f - f_1) \, df_1 df_2 df_3
\nonumber \\
&= \Cooo^2 \iint \Gx(f) \Gx(f_2) \Gy(f_3) \, df_2 df_3\nonumber \\
&= \Cooo^2 P_x P_y \Gx(f).
\end{align}

\subsubsection{The case $\kx = \mx$, $\lx = \kxx$, $\mxx = \lxx$}
We notice that this case is obtained from the above case by swapping
$\kx$ and $\lx$ and will therefore give the same result.

\subsubsection{The total PSD for the second term}
We get the final expression
\begin{align}
\Gxptildeb(f) &= 2 \Cooo^2 P_x P_y \Gx(f).
\end{align}

\subsection{The third term in the PSD}
We now study the third expression, i.e., $\Gxptildec(f) =
\fourier[\Rc(\tau)]$.  Following the above rules for how the indices
can be chosen, we now have two combinations
\begin{align}
\begin{array}{ccc}
\kx = \kxx & \lx = \mx & \mxx = \lxx\\ \kx = \lxx & \lx = \mx & \mxx =
\kxx
\end{array}
\end{align}
The first case is identical to the first case for the second term in
the PSD.  The second case is obtained from the first case by swapping
$\kxx$ and $\lxx$ and will therefore give the same result. We get
\begin{align}
\Gxptildec(f) &= 2 \Cooo^2 P_x P_y \Gx(f).
\end{align}

\subsection{The fourth term in the PSD}
We now study the fourth expression, i.e., $\Gxptilded(f) =
\fourier[\Rd(\tau)]$.  Following the above rules for how the indices
can be chosen, we have two combinations
\begin{align}
\begin{array}{ccc}
\kx = \kxx & \lx = \lxx & \mx = \mxx\\ \kx = \kxx & \lx = \mx & \mxx =
\lxx
\end{array}
\end{align}

\subsubsection{The case $\kx = \kxx$, $\lx = \lxx$, $\mx = \mxx$}
We then have
\begin{align}
\Rd &= \expval \left\{ f_0^3
\sum_{\kx,\lx,\mx} |\Cklm|^2
 e^{i 2 \pi (\kx + \lx - \mx) f_0 \tau} |\xikx|^2 |\xily|^2 |\ximy|^2
 \Gx(\kx f_0) \Gy(\lx f_0) \Gy(\mx f_0) \right\} \nonumber \\
&= f_0^3
\sum_{\kx,\lx,\mx} |\Cklm|^2 e^{i 2 \pi (\kx + \lx - \mx) f_0 \tau}
  \Gx(\kx f_0) \Gy(\lx f_0) \Gy(\mx f_0),
\end{align}
\begin{align}
\Gxptilded(f) &= f_0^3
\sum_{\kx,\lx,\mx} |\Cklm|^2 \Gx(\kx f_0) \Gy(\lx f_0) \Gy(\mx f_0)
\delta(f - (\kx + \lx - \mx)f_0).
\end{align}
Letting $f_0 \to 0$, we get
\begin{align}
\Gxptilded(f) &= \iiint |\mathcal{C}(f_1, f_2, f_3)|^2 \Gx(f_1)
\Gy(f_2) \Gy(f_3) \delta(f - f_1 - f_2 + f_3) \, df_1 df_2 df_3
\nonumber \\
&= \iint |\mathcal{C}(f_1, f_2, f_1 + f_2 - f)|^2 \Gx(f_1) \Gy(f_2)
\Gy(f_1 + f_2 - f) \, df_1 df_2.
\end{align}

\subsubsection{The case $\kx = \kxx$, $\lx = \mx$, $\mxx = \lxx$}
We get
\begin{align}
\Rd &= \expval \left\{ f_0^3
\sum_{\kx,\lx,\lxx} \mathcal{C}_{\kx \lx \lx} \mathcal{C}^*_{\kx \lxx
    \lxx} e^{i 2 \pi \kx f_0 t_1} e^{-i 2 \pi \kx f_0 t_2} |\xikx|^2
    |\xily|^2 |\xilyy|^2 \Gx(\kx f_0) \Gy(\lx f_0) \Gy(\lxx f_0)
    \right\} \nonumber \\
&= f_0^3
\sum_{\kx,\lx,\lxx} \mathcal{C}_{\kx \lx \lx} \mathcal{C}^*_{\kx \lxx
\lxx} e^{i 2 \pi \kx f_0 \tau} \Gx(\kx f_0) \Gy(\lx f_0) \Gy(\lxx f_0),
\end{align}
\begin{align}
\Gxptilded(f) &= f_0^3
\sum_{\kx,\lx,\lxx} \mathcal{C}_{\kx \lx \lx} \mathcal{C}^*_{\kx \lxx
\lxx} \Gx(\kx f_0) \Gy(\lx f_0) \Gy(\lxx f_0) \delta(f - k f_0).
\end{align}
Letting $f_0 \to 0$, we get
\begin{align}
\Gxptilded(f) &= \iiint \mathcal{C}(f_1, f_2, f_2) \mathcal{C}^*(f_1,
f_3, f_3) \Gx(f_1) \Gy(f_2) \Gy(f_3) \delta(f - f_1) \, df_1 df_2 df_3
\nonumber \\
&= \iint \mathcal{C}(f, f_2, f_2) \mathcal{C}^*(f, f_3, f_3) \Gx(f)
\Gy(f_2) \Gy(f_3) \, df_2 df_3 \nonumber \\
&= \Cooo^2 P_y^2 \Gx(f).
\end{align}

\subsubsection{The total PSD for the fourth term}
We get the final expression
\begin{align}
\Gxptildeb(f) &= \iint |\mathcal{C}(f_1, f_2, f_1 + f_2 - f)|^2
\Gx(f_1) \Gy(f_2) \Gy(f_1 + f_2 - f) \, df_1 df_2 + \Cooo^2 P_y^2
\Gx(f).
\end{align}

\subsection{The total PSD}
We can now find the total PSD by summing up all the terms according to
\begin{align}
\label{eq_total_psd}
\Gxptilde &= \Gxptildea + \Gxptildeb + \Gxptildec + \Gxptilded
\nonumber \\
\opt{&= 2 \iint |\mathcal{C}(f_1, f_2, f_1 + f_2 - f)|^2 \Gx(f_1)
\Gx(f_2) \Gx(f_1 + f_2 - f) \, df_1 df_2 + 4 \Cooo^2 P_x^2
\Gx(f)\nonumber \\
&\quad+ 2 \Cooo^2 P_x P_y \Gx(f)\nonumber \\
&\quad+ 2 \Cooo^2 P_x P_y \Gx(f)\nonumber \\
&\quad+ \iint |\mathcal{C}(f_1, f_2, f_1 + f_2 - f)|^2 \Gx(f_1)
\Gy(f_2) \Gy(f_1 + f_2 - f) \, df_1 df_2 + \Cooo^2 P_y^2
\Gx(f)\nonumber \\
&= 2 \iint |\mathcal{C}(f_1, f_2, f_1 + f_2 - f)|^2 \Gx(f_1) \Gx(f_2)
\Gx(f_1 + f_2 - f) \, df_1 df_2 \nonumber \\
&\quad+ \iint |\mathcal{C}(f_1, f_2, f_1 + f_2 - f)|^2 \Gx(f_1)
\Gy(f_2) \Gy(f_1 + f_2 - f) \, df_1 df_2 \nonumber \\
&\quad+ 4 \Cooo^2 P_x^2 \Gx(f) + 2 \Cooo^2 P_x P_y \Gx(f) + 2 \Cooo^2
P_x P_y \Gx(f) + \Cooo^2 P_y^2 \Gx(f)\nonumber \\}
&= 2 \iint |\mathcal{C}(f_1, f_2, f_1 + f_2 - f)|^2 \Gx(f_1) \Gx(f_2)
\Gx(f_1 + f_2 - f) \, df_1 df_2 \nonumber \\
&\quad+ \iint |\mathcal{C}(f_1, f_2, f_1 + f_2 - f)|^2 \Gx(f_1)
\Gy(f_2) \Gy(f_1 + f_2 - f) \, df_1 df_2 \nonumber \\
&\quad+ \Cooo^2 (4 P_x^2 + 4 P_x P_y + P_y^2) \Gx(f).
\end{align}
We notice that the $x$ polarization acting on itself is twice as
effective as the $y$ polarization acting on the $x$ polarization.  As
is clear from the derivation, the reason for this is the level of
degeneracy.  This is similar to the case of the double influence of
XPM as compared to SPM, when the nonlinear Schr\"odinger equation is
approximated by a coupled system with one equation for each WDM
channel~\cite[p.~264]{agrawal_2001_nfo}.  The final term of
(\ref{eq_total_psd}) is of no consequence for the transmission.  The
origin of this term is the phase modulation from the entire
propagating field acting on itself.  In practice, we expect this to
just cause a rotation of the received signal constellations.  However,
in the perturbation analysis, this type of phase modulation gives rise
to these extra terms in the PSD.  As (\ref{eq_total_psd}) can be
considered the main result of this calculation, we repeat the
expression for $\mathcal{C}$ for ease of reference
\begin{align}
\mathcal{C}(f_1, f_2, f_1 + f_2 - f) \equiv \int_{0}^{\Ltot} \gamma(z)
p(z) e^{-i 4 \pi^2 (f_1 - f) (f_2 - f) B(z)} \, dz.
\end{align}
It should be noticed that $\mathcal{C}$ is determined when the
physical parameters of the channel are selected and $|\mathcal{C}|^2$
is a measure of the FWM efficiency.  Then, by choosing the input
signal PSD we obtain the PSD of the NLI.

\section{Transmission system example}
\label{section_example}
We now calculate the resulting expression for a system of the type
considered in~\cite{poggiolini_2011_ptl2}.  The system consists of
$\NSMF$ spans, each containing a standard single-mode fiber (SMF)
followed by an EDFA. There are no dispersion-compensating fibers (DCF)
or any other optical dispersion compensation. Instead, the CD will be
compensated for by using DSP in the receiver.  The first span starts
at $z = z_0 = 0$ and the last span ends at $z = z_\NSMF$.  The fiber
parameters $\alpha$, $\beta_2$, and $\gamma$ are assumed to have no
$z$-dependence.  Furthermore, we assume that identical signals are
launched in the two polarizations, i.e., we set $\Gy = \Gx$ and we
remove the terms that are not due to FWM to rewrite
(\ref{eq_total_psd}) as
\begin{align}
\Gxptilde &= 3 \iint |\mathcal{C}(f_1, f_2, f_1 + f_2 - f)|^2
G_x(f_1) G_x(f_2) G_x(f_1 + f_2 - f) \, df_1
df_2.
\end{align}
We have $B(z) = \beta_2 z$ and temporarily introducing $\kappa = 4
\pi^2 (f_1 - f) (f_2 - f)$ we find
\begin{align}
\mathcal{C}(f_1, f_2, f_1 + f_2 - f) &= \int_{0}^{\Ltot} \gamma p
e^{-i \kappa B(z)} \, dz  \nonumber \\
&= \gamma \sum_{n = 1}^{\NSMF} \int_{z_{n - 1}}^{z_n} e^{-\alpha(z -
z_{n - 1})} e^{-i \kappa \beta_2 z} \, dz \nonumber \\
\opt{&= \gamma
\sum_{n = 1}^{\NSMF} e^{\alpha z_{n - 1}} \int_{z_{n - 1}}^{z_n}
e^{-\alpha z} e^{-i \kappa \beta_2 z} \, dz = \gamma
\sum_{n = 1}^{\NSMF} e^{\alpha z_{n - 1}} \left[ \frac{e^{-\alpha z}
e^{-i \kappa \beta_2 z}}{-\alpha -i \kappa \beta_2} \right]_{z_{n -
1}}^{z_n} \nonumber \\}
\opt{&= -\frac{\gamma}{\alpha + i \kappa \beta_2}
\sum_{n = 1}^{\NSMF} e^{\alpha z_{n - 1}} \left( e^{-\alpha z_n} e^{-i
  \kappa \beta_2 z_n} - e^{-\alpha z_{n - 1}} e^{-i \kappa \beta_2
  z_{n - 1}} \right) \nonumber \\}
&= -\frac{\gamma}{\alpha + i \kappa \beta_2}
\sum_{n = 1}^{\NSMF} \left( e^{-\alpha (z_n - z_{n - 1})} e^{-i \kappa
\beta_2 z_n} - e^{-i \kappa \beta_2 z_{n - 1}} \right) \nonumber \\
&= -\frac{\gamma}{\alpha + i \kappa \beta_2}
\sum_{n = 1}^{\NSMF} \left( e^{-\alpha \LSMF} e^{-i \kappa \beta_2 n
\LSMF} - e^{-i \kappa \beta_2 (n - 1) \LSMF} \right) \nonumber \\
\opt{&= -\frac{\gamma}{\alpha + i \kappa \beta_2}
\sum_{n = 1}^{\NSMF} \left( e^{-\alpha \LSMF} e^{-i \kappa \beta_2
  \LSMF} - 1 \right) e^{-i \kappa \beta_2 (n - 1) \LSMF} \nonumber \\}
\opt{&= -\frac{\gamma}{\alpha + i \kappa \beta_2} ( e^{-\alpha \LSMF} e^{-i
\kappa \beta_2 \LSMF} - 1 ) \frac{1 - e^{-i \kappa \beta_2
\LSMF \NSMF}}{1 - e^{-i \kappa \beta_2 \LSMF}} \nonumber \\}
&= \frac{\gamma}{\alpha + i \kappa \beta_2} ( 1 - e^{-\alpha \LSMF}
e^{-i \kappa \beta_2 \LSMF}) \frac{1 - e^{-i \kappa \beta_2
\LSMF \NSMF}}{1 - e^{-i \kappa \beta_2 \LSMF}},
\end{align}
where we also used the assumption that all SMFs have the same length,
denoted by $\LSMF = z_n - z_{n - 1}$. We find
\begin{align}
|\mathcal{C}(f_1, f_2, f_1 + f_2 - f)|^2 &= \gamma^2 \left| \frac{1 -
  e^{-\alpha \LSMF} e^{-i \kappa \beta_2 \LSMF}}{\alpha + i \kappa
  \beta_2} \right|^2
\left| \frac{1 - e^{-i \kappa \beta_2 \LSMF \NSMF}}{1 - e^{-i \kappa
  \beta_2 \LSMF}} \right|^2 \nonumber \\
\opt{&= \gamma^2 \left| \frac{1 - e^{-\alpha \LSMF} e^{-i \kappa \beta_2
  \LSMF}}{\alpha + i \kappa \beta_2} \right|^2 \frac{2 - 2 \cos
  (\kappa \beta_2 \LSMF \NSMF)}{2 - 2 \cos (\kappa \beta_2 \LSMF)}
  \nonumber \\}
&= \gamma^2 \left| \frac{1 - e^{-\alpha \LSMF} e^{-i \kappa \beta_2
  \LSMF}}{\alpha + i \kappa \beta_2} \right|^2 \frac{\sin^2 (\kappa
  \beta_2 \LSMF \NSMF/2)}{\sin^2 (\kappa \beta_2 \LSMF/2)}.
\end{align}
We then find
\begin{align}
\Gxptilde &= 3 \gamma^2 \iint \left| \frac{1 - e^{-\alpha \LSMF -i 4
\pi^2 (f_1 - f) (f_2 - f) \beta_2 \LSMF}}{\alpha + i 4 \pi^2 (f_1 - f)
(f_2 - f) \beta_2} \right|^2 \frac{\sin^2 [2 \pi^2 (f_1 - f) (f_2 - f)
\beta_2 \LSMF \NSMF]}{\sin^2 [2 \pi^2 (f_1 - f) (f_2 - f) \beta_2
\LSMF]} \nonumber \\
& \quad \times G_x(f_1) G_x(f_2) G_x(f_1 + f_2
- f) \, df_1 df_2.
\end{align}
In order to compare this expression with the coherent expression
in~\cite[Eq.~(18)]{carena_2012_jlt}, we must account for the fact that
a different convention for the Manakov equation is used
in~\cite{carena_2012_jlt}. Thus, we need to replace $\gamma$ with $8
\gamma/9$. Furthermore, the NLI PSD
in~\cite[Eq.~(18)]{carena_2012_jlt} is the summation over both
polarizations, i.e., we here get $G_\text{NLI} = \Gxptilde + \Gyptilde = 2 \Gxptilde$.
Similarly, we here have $G_\text{Tx} = G_x + G_y = 2
G_x$. (From~\cite[Eq.~(13)]{carena_2012_jlt}, it seems like
$G_\text{Tx}$ is equal to $G_x$. However, this is not the intended
meaning of $G_\text{Tx}$. We thank the authors
of~\cite{carena_2012_jlt} for explaining this.)  Finally, there is a
difference in how the attenuation is defined.  For the definition used
here, we refer to (\ref{eq_manakov}) and (\ref{eq_alpha_def}).  When
these differences in notation are taken into account, the results
become identical.  \opt{
\newpage
\begin{align}
\frac{G_\text{NLI}}{2} &= 3 (8 \gamma/9)^2 \iint \left| \frac{1 -
  e^{-2 \alpha \LSMF -i 4 \pi^2 (f_1 - f) (f_2 - f) \beta_2 \LSMF}}{2
  \alpha + i 4 \pi^2 (f_1 - f) (f_2 - f) \beta_2} \right|^2
\frac{\sin^2 [2 \pi^2 (f_1 - f) (f_2 - f) \beta_2 \LSMF \NSMF]}{\sin^2
  [2 \pi^2 (f_1 - f) (f_2 - f) \beta_2 \LSMF]} \nonumber \\
& \quad \times \frac{G_\text{Tx}(f_1)}{2} \frac{G_\text{Tx}(f_2)}{2}
  \frac{G_\text{Tx}(f_1 + f_2 - f)}{2} \, df_1 df_2.
\end{align}
}

\opt{
\begin{align}
G_\text{NLI} &= \frac{3}{4} \frac{64}{81} \gamma^2 \iint \left|
  \frac{1 - e^{-2 \alpha \LSMF -i 4 \pi^2 (f_1 - f) (f_2 - f) \beta_2
  \LSMF}}{2 \alpha + i 4 \pi^2 (f_1 - f) (f_2 - f) \beta_2} \right|^2
  \frac{\sin^2 [2 \pi^2 (f_1 - f) (f_2 - f) \beta_2 \LSMF
  \NSMF]}{\sin^2 [2 \pi^2 (f_1 - f) (f_2 - f) \beta_2 \LSMF]}
  \nonumber \\
& \quad \times G_\text{Tx}(f_1) G_\text{Tx}(f_2) G_\text{Tx}(f_1 + f_2
  - f) \, df_1 df_2 \nonumber \\
&= \frac{16}{27} \gamma^2 \iint \left|
  \frac{1 - e^{-2 \alpha \LSMF -i 4 \pi^2 (f_1 - f) (f_2 - f) \beta_2
  \LSMF}}{2 \alpha + i 4 \pi^2 (f_1 - f) (f_2 - f) \beta_2} \right|^2
  \frac{\sin^2 [2 \pi^2 (f_1 - f) (f_2 - f) \beta_2 \LSMF
  \NSMF]}{\sin^2 [2 \pi^2 (f_1 - f) (f_2 - f) \beta_2 \LSMF]}
  \nonumber \\
& \quad \times G_\text{Tx}(f_1) G_\text{Tx}(f_2) G_\text{Tx}(f_1 + f_2
  - f) \, df_1 df_2
\end{align}
}

\section{Conclusion}
We have presented a derivation of the power spectral density of the
nonlinear interference under the assumptions that (i)~the nonlinear
effects are weak, (ii)~the signal can be written as suggested
in~\cite{poggiolini_2011_ptl2}, and (iii)~the signal--noise
interaction can be neglected.  Using these three assumptions, we
obtain the PSD expression for a general system.  Applying the result
to a system consisting entirely of SMF and accounting for the
differences in notation, we obtain a result identical to that
presented in~\cite{poggiolini_2011_ptl2,
poggiolini_2011_icton,carena_2012_jlt}.


\begin{thebibliography}{10}
\providecommand{\url}[1]{#1}
\csname url@samestyle\endcsname
\providecommand{\newblock}{\relax}
\providecommand{\bibinfo}[2]{#2}
\providecommand{\BIBentrySTDinterwordspacing}{\spaceskip=0pt\relax}
\providecommand{\BIBentryALTinterwordstretchfactor}{4}
\providecommand{\BIBentryALTinterwordspacing}{\spaceskip=\fontdimen2\font plus
\BIBentryALTinterwordstretchfactor\fontdimen3\font minus
  \fontdimen4\font\relax}
\providecommand{\BIBforeignlanguage}[2]{{%
\expandafter\ifx\csname l@#1\endcsname\relax
\typeout{** WARNING: IEEEtran.bst: No hyphenation pattern has been}%
\typeout{** loaded for the language `#1'. Using the pattern for}%
\typeout{** the default language instead.}%
\else
\language=\csname l@#1\endcsname
\fi
#2}}
\providecommand{\BIBdecl}{\relax}
\BIBdecl

\bibitem{carena_2010_ecoc}
A.~Carena, G.~Bosco, V.~Curri, P.~Poggiolini, M.~{Tapia Taiba}, and
  F.~Forghieri, ``Statistical characterization of {PM-QPSK} signals after
  propagation in uncompensated fiber links,'' in \emph{European Conference on
  Optical Communication (ECOC)}, 2010, p. P4.07.

\bibitem{poggiolini_2011_ptl2}
P.~Poggiolini, A.~Carena, V.~Curri, G.~Bosco, and F.~Forghieri, ``Analytical
  modeling of nonlinear propagation in uncompensated optical transmission
  links,'' \emph{IEEE Photon.\ Technol.\ Lett.}, vol.~23, no.~11, pp. 742--744,
  June 2011.

\bibitem{poggiolini_2011_icton}
P.~Poggiolini, G.~Bosco, A.~Carena, V.~Curri, and F.~Forghieri, ``A simple and
  accurate model for non-linear propagation effects in uncompensated coherent
  transmission links,'' in \emph{International Conference on Transparent
  Optical Networks (ICTON)}, 2011, p. We.B1.3.

\bibitem{carena_2012_jlt}
A.~Carena, V.~Curri, G.~Bosco, P.~Poggiolini, and F.~Forghieri, ``Modeling of
  the impact of non-linear propagation effects in uncompensated optical
  coherent transmission links,'' \emph{J.\ Lightw.\ Technol.}, vol.~30, no.~10,
  pp. 1524--1539, May 2012.

\bibitem{chen_2010_oe}
X.~Chen and W.~Shieh, ``Closed-form expressions for nonlinear transmission
  performance of densely spaced coherent optical {OFDM} systems,'' \emph{Opt.\
  Express}, vol.~18, no.~18, pp. 19\,039--19\,054, Aug. 2010.

\bibitem{inoue_1995_jlt}
K.~Inoue and H.~Toba, ``Fiber four-wave mixing in multi-amplifier systems with
  nonuniform chromatic dispersion,'' \emph{J.\ Lightw.\ Technol.}, vol.~13,
  no.~1, pp. 88--93, Jan. 1995.

\bibitem{mecozzi_2000_ptl}
A.~Mecozzi, C.~B. Clausen, and M.~Shtaif, ``Analysis of intrachannel nonlinear
  effects in highly dispersed optical pulse transmission,'' \emph{IEEE Photon.\
  Technol.\ Lett.}, vol.~12, no.~4, pp. 392--394, Apr. 2000.

\bibitem{ablowitz_2000_ol}
M.~J. Ablowitz and T.~Hirooka, ``Resonant nonlinear intrachannel interactions
  in strongly dispersion-managed transmission systems,'' \emph{Opt.\ Lett.},
  vol.~25, no.~24, pp. 1750--1752, Dec. 2000.

\bibitem{johannisson_2001_ol}
P.~Johannisson, D.~Anderson, A.~Berntson, and J.~M{\aa}rtensson, ``Generation
  and dynamics of ghost pulses in strongly dispersion-managed fiber-optic
  communication systems,'' \emph{Opt.\ Lett.}, vol.~26, no.~16, pp. 1227--1229,
  Aug. 2001.

\bibitem{bononi_2008_crp}
A.~Bononi, P.~Serena, and M.~Bertolini, ``Unified analysis of weakly-nonlinear
  dispersion-managed optical transmission systems using a perturbative
  approach,'' \emph{Comptes Rendus Physique}, vol.~9, pp. 947--962, Nov. 2008.

\bibitem{bononi_2012_oe}
A.~Bononi, P.~Serena, N.~Rossi, E.~Grellier, and F.~Vacondio, ``Modeling
  nonlinearity in coherent transmissions with dominant
  intrachannel-four-wave-mixing,'' \emph{Opt.\ Express}, vol.~20, no.~7, pp.
  7777--7791, Mar. 2012.

\bibitem{yan_2011_ecoc}
W.~Yan, Z.~Tao, L.~Dou, L.~Li, S.~Oda, T.~Tanimura, T.~Hoshida, and J.~C.
  Rasmussen, ``Low complexity digital perturbation back-propagation,'' in
  \emph{European Conference on Optical Communication (ECOC)}, 2011, p.
  Tu.3.A.2.

\bibitem{wang_1999_jlt}
D.~Wang and C.~R. Menyuk, ``Polarization evolution due to the {Kerr}
  nonlinearity and chromatic dispersion,'' \emph{J.\ Lightw.\ Technol.},
  vol.~17, no.~12, pp. 2520--2529, Dec. 1999.

\bibitem{proakis_dig_comm_5ed}
J.~G. Proakis and M.~Salehi, \emph{Digital Communications}, 5th~ed.\hskip 1em
  plus 0.5em minus 0.4em\relax McGraw-Hill, 2008.

\bibitem{agrawal_2001_nfo}
G.~P. Agrawal, \emph{Nonlinear Fiber Optics}, 3rd~ed.\hskip 1em plus 0.5em
  minus 0.4em\relax Academic Press, 2001.

\end{thebibliography}
\end{document}